\documentclass[aps,prb,twocolumn,longbibliography]{revtex4-2}

\usepackage{graphicx}
\usepackage{amsmath}
\usepackage{physics}
\usepackage{braket}
\usepackage{amssymb}
\usepackage{mathtools}
\usepackage{multirow}
\usepackage{setspace}
\usepackage[colorlinks,citecolor=blue,urlcolor=black]{hyperref}

\def\ben{\begin{equation}}
\def\een{\end{equation}}

\begin{document}

\title{Quasiparticle Electronic Structure of Two-Dimensional Heterotriangulene-Based Covalent Organic Frameworks Adsorbed on Au(111)}

\author{Joseph Frimpong}
\author{Zhen-Fei Liu}
\email{zfliu@wayne.edu}
\affiliation{Department of Chemistry, Wayne State University, Detroit, MI 48202, USA}

\date{\today}

\begin{abstract}
The modular nature and unique electronic properties of two-dimensional (2D) covalent organic frameworks (COFs) make them an attractive option for applications in catalysis, optoelectronics, and spintronics. The fabrications of such devices often involve interfaces formed between COFs and substrates. In this work, we employ the first-principles $GW$ approach to accurately determine the quasiparticle electronic structure of three 2D carbonyl bridged heterotriangulene-based COFs featuring kagome lattice, with their properties ranging from a semi-metal to a wide-gap semiconductor. Moreover, we study the adsorption of these COFs on Au(111) surface and characterize the quasiparticle electronic structure at the heterogeneous COF/Au(111) interfaces. To reduce the computational cost, we apply the recently developed dielectric embedding $GW$ approach and show that our results agree with existing experimental measurement on the interfacial energy level alignment. Our calculations illustrate how the many-body dielectric screening at the interface modulates the energies and shapes of the kagome bands, the effective masses of semiconducting COFs, as well as the Fermi velocity of the semi-metallic COF. 
\end{abstract}

\maketitle

\section{Introduction}
Covalent organic frameworks (COFs) are micro-porous materials in which non-metallic light atoms form periodic structures via covalent bonds \cite{DY17,YOOC03}. These crystalline materials could feature different types of bond connectivities, giving rise to a wide range of two-dimensional (2D) \cite{LCXX19,CBOO05,ZXWP14,DAJS15} and three-dimensional (3D) \cite{UHFK09,EHMC07,LDYW16,KHYC20, GCFQ20} topologies. COFs have attracted great attention because of their diverse properties such as high porosity \cite{LZCL18,LJAZ20, EHMC07}, high charge carrier mobilities \cite{WGAF11}, modularity in synthesis \cite{FWLY20,W20,PZLQ17}, etc., in addition to the non-trivial topological properties that are intrinsic to the lattice symmetry \cite{SLKH20,GHLD20}. This makes them excellent candidates for applications in gas sensing \cite{LHLZ19}, energy storage \cite{XJZW15, LJLL20}, spintronics \cite{LTWZ14,JHL19}, optoelectronics \cite{MMB17}, and catalysis \cite{LDZK15, GOSA16, DGWZ11}. While the synthesis of perfectly crystalline COFs remains a challenge \cite{HL20}, most 2D COFs are synthesized by Ullmann coupling \cite{SGAY17, ZPAC08, BTCA09} and self-assembled on metal or insulating substrates \cite{SGAY17,KC20,MBSS16}. Moreover, their practical applications in field effect transistor-type devices often involve interfaces between COFs and metal substrates \cite{KC19,MLJS15}. Therefore, it is of paramount significance to understand how properties of COFs are modulated by the substrates on which the COFs are synthesized or assembled. 

First-principles calculations are an indispensable tool in understanding the properties of COFs and their interactions with metal substrates. For the latter, a key physics is the many-body interaction between the substrate and the COF, namely the dielectric screening of the Coulomb interaction within the COF due to the substrate. This effect renormalizes the fundamental gap of the COF adsorbed on a metal substrate compared to the freestanding COF and is known as the ``image-charge effect" \cite{I71,LK73,NHL06,TR09}.  Due to the typical large size of the COFs and their interfaces with substrates, most theoretical works have employed density functional theory (DFT) \cite{HK64,KS65} for the determination of the structural, electronic, and magnetic properties \cite{YDHF15, GG17,JH19, SGAY17,FCC19, LTWZ14, TLZM19, RDBV20, MBSS16, ZBB20}. However, while most functionals can predict the total energy and density related properties to a good accuracy, the orbital energy levels and energy level alignments of COF-metal interfaces are generally beyond the reach of many functionals. This is because the latter are  quasiparticle properties that require a many-body treatment of the correlation effects \cite{NHL06,TR09}. Hybrid functionals such as HSE (Heyd–Scuseria–Ernzerhof) \cite{HSE03} improve on band gaps of bulk semiconductors over local and semi-local functionals \cite{HPS11} and have been applied to study COFs and their adsorption on substrates \cite{JWZ15,R16,SGAY17,JH19,YMMG20,CZJZ18,LTWZ14}. However, they are often still not quantitatively accurate in the determination of the energy level alignment \cite{BTNK11} at heterogeneous interfaces. The development of hybrid functionals for this purpose is still at its infancy \cite{LERK17}.

Compared to DFT, many-body perturbation theory (MBPT) provides a rigorous theoretical framework in computing quasiparticle properties \cite{ORR02}. Its common approximation is the $GW$ approach \cite{H65,HL86} where $G$ is the Green's function and $W$ is the screened Coulomb interaction. However, the relatively high computational cost of first-principles $GW$ has hindered its routine applications to large systems, such as the COF-substrate interfaces. Simplified $GW$-based schemes exist, such as the DFT+$\Sigma$~\cite{QVCL07,ELNK15,NHL06}. However, DFT+$\Sigma$~cannot capture the dynamical effects of the self-energy and requires an explicit specification of the image-charge plane \cite{ELNK15}, which may be ambiguous in certain cases. Moreover, it is not trivial to use the original version of DFT+$\Sigma$~to treat a \emph{periodic} adsorbate such as the COF, because it would be a very drastic approximation to model an orbital of the periodic adsorbate using a single point charge. In fact, Ref. \cite{LZM15} generalized the original DFT+$\Sigma$~idea in a model to study the properties of COF-substrate interfaces. Although direct $GW$ calculations have been employed in studying freestanding COFs \cite{LZM15,ZM12,WQ20}, to the best of our knowledge, we have not found $GW$-based calculations of COF-substrate interfaces. 

One bottleneck of large-scale $GW$ calculations is the high computational cost of the non-interacting Kohn-Sham (KS) polarizability in the random-phase approximation \cite{DSSJ12}, $\chi^0$, which scales as $\mathcal{O}(N^4)$ with $N$ being the system size. An emerging approach for large-scale interfaces leverages the (approximate) additivity of the substrate and adsorbate KS polarizabilities \cite{UBSJ14,LJLN19,XCQ19}, greatly reducing the computational cost. This approximation holds when the substrate-adsorbate hybridization approaches zero, valid for most physisorptions. Even with these approaches, one often still needs to compute many empty bands for the combined COF-substrate interface in order to converge the self-energy. For large simulation cells, this is still a formidable task. To tackle this challenge, we proposed an approach in Ref. \cite{L20} where the explicit $GW$ calculations are confined to a simulation cell that only contains the adsorbate, which is embedded in the dielectric environment of the substrate. This is termed ``dielectric embedding $GW$" and was shown to be accurate for a few prototypical physisorbed interfaces. 

In this work, we apply the dielectric embedding $GW$ approach developed in Ref. \cite{L20} to a series of 2D heterotriangulene-based COFs adsorbed on Au(111) surface. These COFs feature kagome lattice and we focus on how their electronic properties are modulated by the many-body screening of the Au(111) substrate. The substrate effect is elucidated via a comparison with $GW$ calculations of the freestanding COFs. Note that Ref. \cite{SGAY17} studied the carbonyl-bridged triphenylamine(CTPA)/Au(111) interface using both scanning tunneling spectroscopy and HSE calculations. Our result on the CTPA/Au(111) is in quantitative agreement with the experimental measurement \cite{SGAY17}, validating the method we use. Our results for other COF/Au(111) systems provide a theoretical benchmark for future experiments.

This paper is structured as follows: we discuss the structure of the COFs and the COF/Au(111) systems in Sec. \ref{structure}, and detail the methodology used for freestanding COFs in Sec. \ref{method_COF} and that for the interfaces in Sec. \ref{method_composite}. We then discuss our results of the various COFs and their respective interfaces with Au(111) in Sec. \ref{results} and conclude in Sec. \ref{conclusion}.   

\section{Systems and Methodology}
\subsection{COF systems and geometry relaxations} \label{structure}

\begin{figure}[hbtp]
\centering
\includegraphics[width=3.3in]{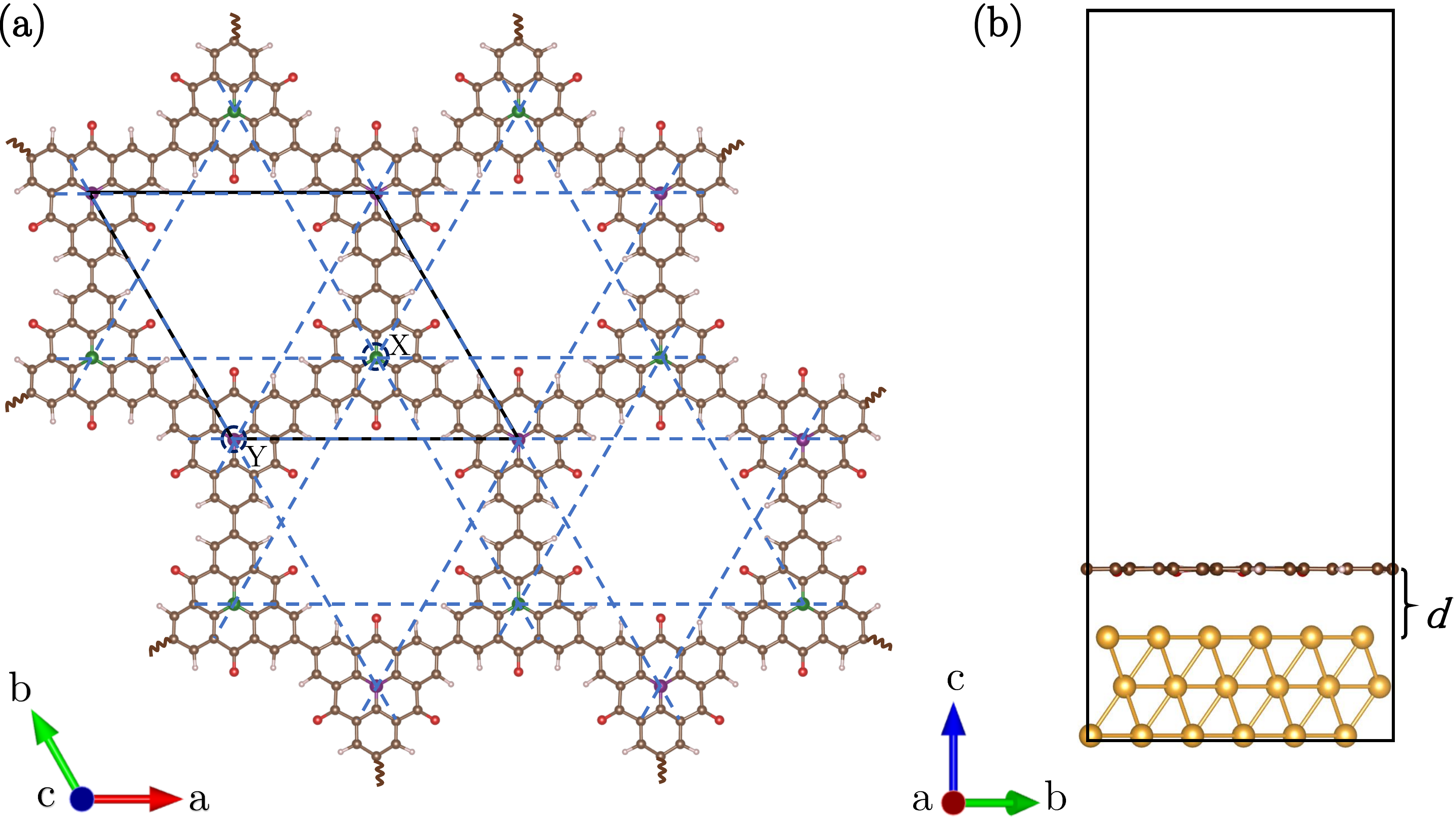}
\caption{(a) COF structures studied in this work. X (green) and Y (purple) are two sites hosting different or same atoms. CTPA: X=N and Y=N; CTP: X=C and Y=C; CTP-BN: X=B and Y=N. The black lines represent the hexagonal simulation cell and the blue dashed lines represent the kagome lattice; (b) A side view of the COF/Au(111) interface where we denote the adsorption height by $d$. The black lines represent the simulation cell. This figure is rendered using VESTA \cite{MI11}. Color code: H - pink; C - brown; O - red; Au - gold.}
\label{fig:cof}
\end{figure}

2D heterotriangulene-based COFs possess an Archimedean net topology forming the so-called ``kagome lattice", which features corner-sharing triangular moieties and a hexagonal unit cell \cite{SLKH20} [Figure \ref{fig:cof}(a)]. A well-studied COF with kagome lattice is the CTPA which has a nitrogen atom at the center of each triangulene [X=N and Y=N in Figure \ref{fig:cof}(a)]. By replacing all nitrogen atoms with carbon, one obtains the structure of the triangulene polymer (CTP) [X=C and Y=C in Figure \ref{fig:cof}(a)]. CTPA has been studied both experimentally and theoretically \cite{SGAY17,JH19,GMHM20}, while CTP has only been studied by theory \cite{JH19}. Here, in addition to CTPA and CTP, we computationally explore another structure by replacing the nitrogen atoms in half of the triangulenes by boron. We call the resulting structure CTP-BN [X=B and Y=N in Figure \ref{fig:cof}(a)]. The relation between CTP and CTP-BN is therefore similar to that between graphene and hexagonal boron nitride (hBN). Variable-cell relaxations of these COFs are performed until all residual forces are less than 0.05 eV/\AA, in simulation cells where symmetries are fixed as hexagonal and the lattice constant along the $c$ direction is fixed as 30 \AA. In such calculations, we use the vdW-DF-cx \cite{BH14} functional as implemented in the Quantum ESPRESSO \cite{GABB17} package, the optimized norm-conserving Vanderbilt (ONCV) pseudopotentials \cite{H13}, a 100 Ry kinetic energy cutoff, and the $\Gamma$-point sampling. The relaxed in-plane lattice parameters of CTPA, CTP, and CTP-BN are 17.36 \AA, 17.41 \AA, and 17.46 \AA, respectively. 

We then model the COF/Au(111) interfaces in simulation cells of hexagonal symmetry with 30 \AA~as the lattice constant along the $c$ direction, containing three layers of Au(111), as shown in Figure \ref{fig:cof}(b). We fix the in-plane lattice parameters of all COF/Au(111) simulation cells to be 17.36 \AA, the optimized value for the freestanding CTPA. We use the same in-plane lattice parameters for all interface systems so that we can use the same $\chi^0$ from the Au(111) unit cell to more than one interfaces (to be specific, CTPA and CTP-BN) in our dielectric embedding $GW$ calculations. This in-plane lattice parameter corresponds to $6\times6$ Au atoms in the (111) plane with an equivalent Au face-centered cubic (FCC) lattice constant of 4.092 \AA. This value is very similar to both the experimental lattice constant of 4.065 \AA~\cite{D25,RLT16} and the optimal vdW-DF-cx lattice constant of 4.108 \AA~\cite{AS16}. Therefore, the strain on the Au atoms is very small.  

The COF/Au(111) interfaces are relaxed with fixed lattice parameters and all 108 Au atoms in the substrate fixed in their bulk positions. All atoms in the COF are allowed to relax until all forces are below 0.05 eV/\AA. We use a kinetic energy cutoff of 70 Ry and a 3$\times$3$\times$1 $\mathbf{k}$-mesh, and the vdW-DF-cx \cite{BH14} functional. The average adsorption height [$d$ in Figure \ref{fig:cof}(b)] for CTPA, CTP, and CTP-BN on the Au(111) substrate is 3.28 \AA, 3.04 \AA, and 3.16 \AA, respectively. Since the COFs are not perfectly flat, we calculate this average adsorption height by averaging out the adsorption heights of all atoms in the COF. Within each COF, the largest deviation from the average values is 0.03 \AA~for both CTPA and CTP, and 0.15 \AA~for CTP-BN. 

\subsection{Quasiparticle electronic structure of freestanding COFs} \label{method_COF}
For the purpose of comparing the electronic structure of the freestanding COFs and those adsorbed on Au substrate, we calculate the quasiparticle properties of all freestanding COFs in simulation cells whose in-plane lattice parameters are consistent with those of the interface, i.e., 17.36 \AA, instead of the individually optimized value for each COF. The size of the cells along the $c$ direction is fixed to be 10 \AA, a value that we used before \cite{LJLN19,L20} and know works well for flat molecules. The COF coordinates are fixed as those in the relaxed COF/Au(111) interfaces. We perform perturbative $G_{0}W_{0}$ calculations as implemented in the BerkeleyGW package \cite{DSSJ12}, using the Perdew-Burke-Ernzerhof (PBE) functional \cite{PBE96} as the mean-field starting point. The PBE calculations of the freestanding CTPA and CTP-BN are preformed using a kinetic energy cutoff of 70 Ry and a $3\times3\times1$ $\mathbf{k}$-mesh. For CTP, we use a much higher $\mathbf{k}$-mesh of $6\times6\times1$ to obtain a more accurate description of the bands near the Dirac point. 

The $GW$ calculation uses a $\mathbf{q}$-mesh that is the same as the $\mathbf{k}$-mesh in the corresponding PBE calculation. We have checked that a kinetic energy cutoff of 7.2 Ry in the dielectric function - corresponding to 6000 bands in the aforementioned simulation cell - converges the self-energies of individual levels (although the gaps converge at a lower cutoff, 3.3 Ry, corresponding to 2000 bands). To correctly model the $\mathbf{q} \rightarrow 0$ limit in the dielectric function, for CTPA and CTP-BN, we include 200 bands in the wavefunctions calculated on a shifted $\mathbf{k}$-grid. For CTP, where bands with linear dispersion touch at the Dirac point (see below), we include 200 bands in the wavefunctions calculated in a much finer $\mathbf{k}$-mesh, $12\times12\times1$. In all our calculations, we use a slab truncation scheme \cite{I06} to eliminate the spurious Coulomb interactions along the $c$ direction. In the self-energy calculations, we model the frequency dependence of the self-energy using the Hybertson-Louie generalized plasmon pole model \cite{HL86}. The semiconductor screening is applied to CTPA and CTP-BN, and the graphene screening is applied to CTP. The static reminder \cite{DSJC13} is applied to all systems to facilitate the convergence with respect to the number of bands.

\subsection{Dielectric embedding $GW$ for COF/Au(111) interfaces} \label{method_composite}
\begin{figure}[h]
\includegraphics[width=3.3in]{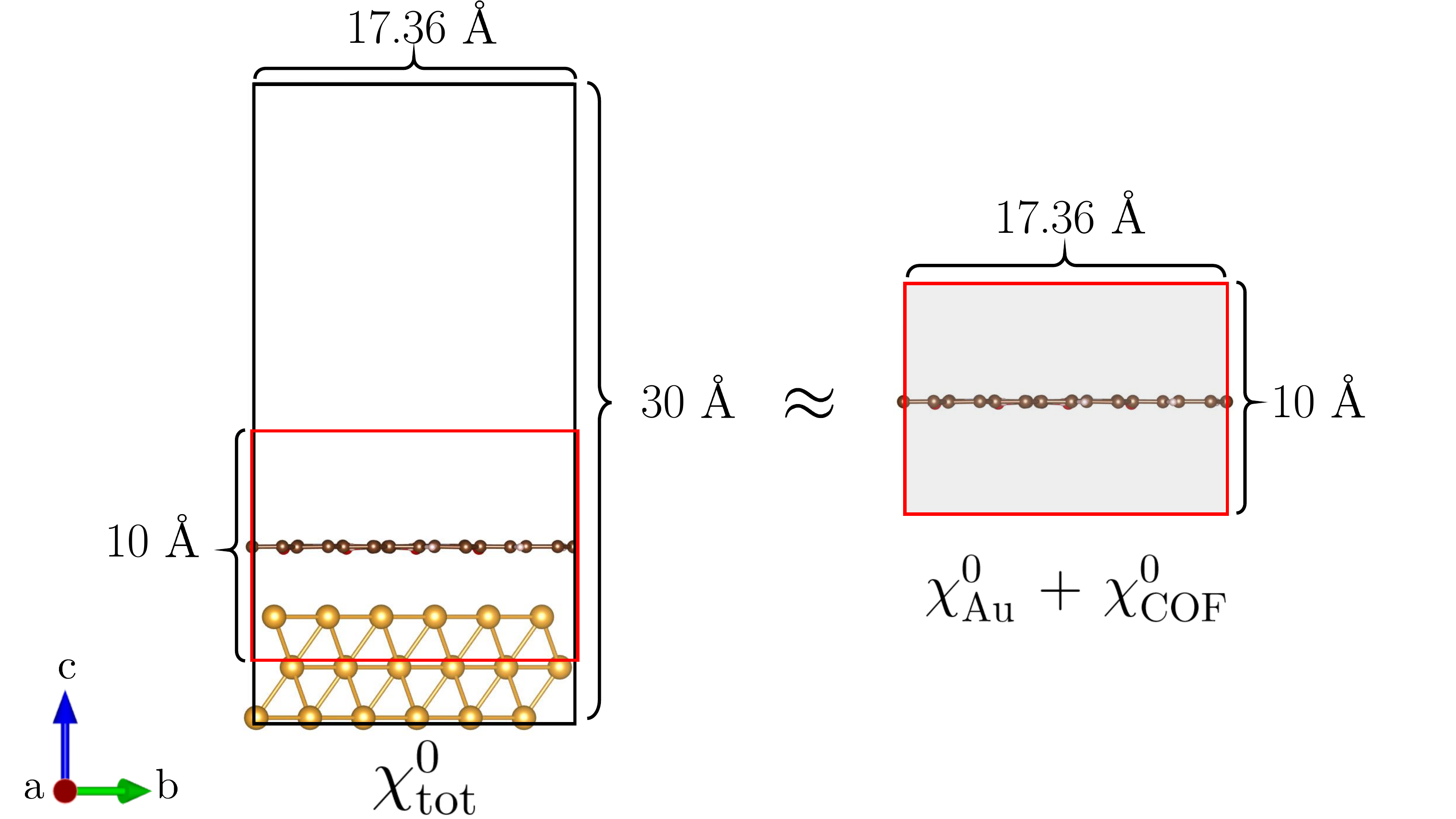}
\caption{A schematic view of the dielectric embedding $GW$ approach applied to the COF/Au(111) interface. Instead of computing the $\chi^0_{\rm tot}$ and self-energies in the interface (left panel), the $GW$ calculations are confined to a smaller simulation cell of COF only (right panel), with the same dimensions along $a$ and $b$ directions as the interface, but a much smaller lattice constant along $c$. The gray background in the right panel indicates that the COF is embedded in the dielectric environment of the substrate, truncated into the red box as illustrated on the left panel.}
\label{fig:embeddingGW_method}
\end{figure}

Given the size of the COF/Au(111) interfaces, it is challenging to perform direct $GW$ calculations. For such systems, we apply the dielectric embedding $GW$ approach developed in Ref. \cite{L20}. Physically, this method approximates the substrate as a dielectric environment in which the adsorbate is embedded, as schematically shown in Fig. \ref{fig:embeddingGW_method}. We refer the interested readers to Ref. \cite{L20} for details of this method, and only list the major steps and computational parameters here. Overall, we approximate the KS polarizability of the interface, $\chi^0_{\rm tot}$, as a sum of the KS polarizabilities of the Au substrate ($\chi^0_{\rm Au}$) and that of the COF ($\chi^0_{\rm COF}$), i.e., $\chi_{\rm tot}^0\approx\chi^0_{\rm Au}+\chi^0_{\rm COF}$. The calculation of $\chi^0_{\rm COF}$ was detailed in Sec. \ref{method_COF}. For the $\chi^0_{\rm Au}$, we first calculate this quantity in the Au(111) unit cell and then uses reciprocal-space folding and real-space truncation techniques to obtain its corresponding quantity within the COF simulation cell, whose dimension is consistent with $\chi^0_{\rm COF}$. In particular, we truncate the $\chi^0_{\rm Au}$ into a region of the same size as the freestanding COF simulation cell and centered at the COF adsorbate on Au(111), as illustrated using the red box in Fig. \ref{fig:embeddingGW_method}.

After the $\chi^0_{\rm tot}$ is summed up from its constituents in the COF simulation cell, the dielectric function is calculated using $\epsilon^{-1}_{\rm tot}= [1 - v\chi^0_{\rm tot}]^{-1}$ where $v$ is the Coulomb interaction, and the self-energy is calculated via $\braket{\phi_i^{\rm COF}|\Sigma[G^{\rm COF}\tilde{W}^{\rm tot}]|\phi_i^{\rm COF}}$. In this equation, $\ket{\phi_i^{\rm COF}}$ is a COF orbital of interest, e.g., the valance band maximum (VBM) or the conduction band minimum (CBM) or any other bands. $\tilde{W}^{\rm tot}=\epsilon^{-1}_{\rm tot}v$ is the screened Coulomb interaction evaluated in the COF simulated cell. 

Here, we provide further details regarding the calculation of $\chi^0_{\rm Au}$ in the Au(111) unit cell. The hexagonal Au(111) simulation cell consists of 3 Au atoms with 1 on each layer, and has a lattice constant of 2.893 \AA~along $a$ and $b$ directions (1/6 of that in the interface) and 30 \AA~along the $c$ direction (same as that in the interface). For the Au(111) unit cell used for the embedding $GW$ calculations of CTPA and CTP-BN, we use an $18\times18\times1$ $\mathbf{k}$-mesh and 501 bands for the wavefunction needed for $\chi^{0}_{\rm Au}(\mathbf{q}\neq0)$, and a $36\times36\times1$ and 180 bands for the wavefunction needed for $\chi^{0}_{\rm Au}(\mathbf{q} \rightarrow 0)$. For the Au(111) unit cell used for the embedding $GW$ calculation of CTP, we use the same number of bands but a finer $\mathbf{k}$-mesh: $36\times36\times1$ and $72\times72\times1$ for the wavefunctions needed for $\chi^{0}_{\rm Au}(\mathbf{q}\neq0)$ and $\chi^{0}_{\rm Au}(\mathbf{q} \rightarrow 0)$, respectively. All the above $\mathbf{k}$-mesh values are chosen to be commensurate with the $\mathbf{k}$-mesh of the corresponding COFs. Similarly, a kinetic energy cutoff of 7.2 Ry is used for $\chi^{0}_{\rm Au}$, consistent with that of $\chi^{0}_{\rm COF}$. 

\section{Results and Discussion}\label{results}
 
\subsection{CTPA and its adsorption on Au(111)}
\begin{figure}[htp]
\includegraphics[width=3.3in]{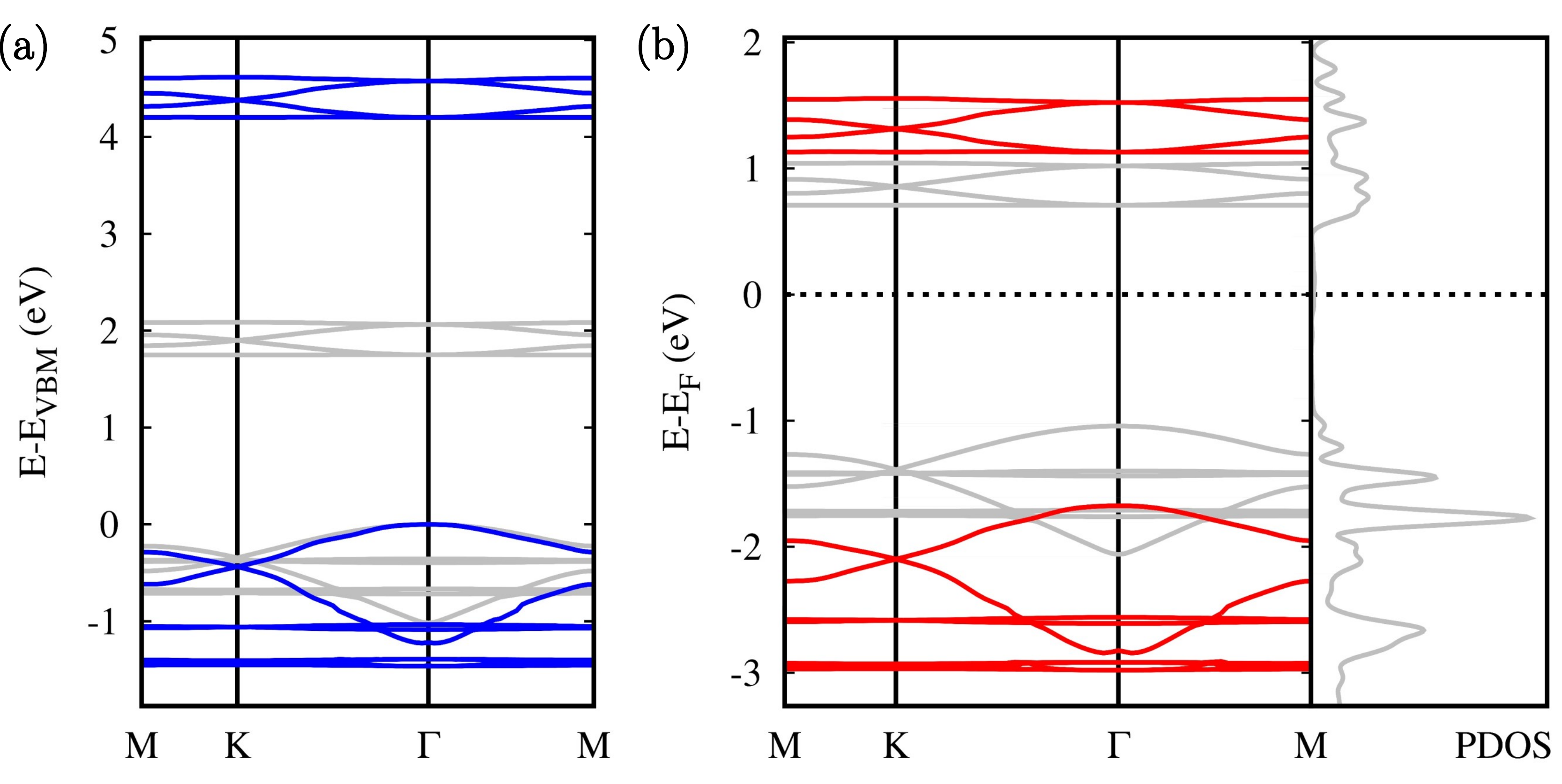}
\caption{(a) PBE (gray) and $GW$ (blue) band structures of the freestanding CTPA; (b) PBE (gray) and dielectric embedding $GW$ (red) band structures of CTPA, with band energies shifted such that the value of PBE VBM matches the energy level alignment at the CTPA/Au(111) interface, i.e., $E_{\rm VBM}-E_{\rm F}$. The PBE PDOS of the CTPA/Au(111) interface onto the CTPA is shown for comparison. The positions of the embedding $GW$ bands are chosen such that their energy differences from the corresponding PBE bands match the self-energies in the dielectric embedding $GW$ calculation. In this way, the dielectric embedding $GW$ VBM (CBM) reflects the prediction of $E_{\rm VBM}-E_{\rm F}$ ($E_{\rm CBM}-E_{\rm F}$) at the CTPA/Au(111) interface using this approach.}
\label{fig:CTPA_pdos_bandstructure}
\end{figure}

The electronic structure of the CTPA/Au(111) interface has been characterized both experimentally \cite{SGAY17} and theoretically \cite{JH19,SGAY17}, which provides a benchmark for our dielectric embedding calculations. Figure \ref{fig:CTPA_pdos_bandstructure}(a) shows the band structure of the freestanding CTPA calculated in PBE (gray) and $GW$ (blue), respectively. For comparison, we align the VBM of the PBE and $GW$ band structures at $E=0$. PBE and $GW$ share the same qualitative features of the kagome bands: for conduction bands, two dispersive Dirac bands cross at the K point and are sandwiched between two flat bands. The CBM is at $\Gamma$, whose energy is nearly degenerate with other $\mathbf{k}$ points. For valance bands, both PBE and $GW$ predict the VBM at $\Gamma$, giving rise to a direct fundamental gap, calculated to be 1.75 eV in PBE and 4.21 eV in $GW$. In the $GW$ results, the top valance band appears to be slightly more dispersive than that in PBE (see the effective mass results below). Additionally, the six occupied flat bands (with two pairs of degenerate bands) are lower in energy than the two Dirac bands in $GW$ (i.e., below -1 eV), while they are energetically between the two Dirac bands in PBE (i.e., between 0 and -1 eV). Note that this difference in the relative positions between the flat and the Dirac bands is also captured by the HSE functional \cite{SGAY17}, and is due to the different localizations of these bands. To be specific, the two Dirac bands consist of delocalized $\pi$ orbitals, and the flat bands are mainly localized on the oxygen atoms or nearby. Our findings here are consistent with the fact that the magnitude of the $GW$ self-energy generally depends on the localization of an orbital and the mean-field starting point \cite{MRMC11}. 

Figure \ref{fig:CTPA_pdos_bandstructure}(b) shows the electronic structure at the CTPA/Au(111) interface. We first show the PBE projected density of states (PDOS) of the interface onto the CTPA, with $E=0$ defined as the Fermi level ($E_{\rm F}$) of the interface. Then we show the PBE band structure (gray) of the freestanding CTPA [the same one as in Figure \ref{fig:CTPA_pdos_bandstructure}(a)] again, but shift the energies such that the PBE VBM matches $E_{\rm VBM}-E_{\rm F}$ of the CTPA/Au(111) interface. One can then correlate the band structure with the PDOS. Due to the weak coupling nature of the interface and the neglect of band gap renormalization in PBE, the PBE CBM-VBM gap of the freestanding CTPA matches the CBM-VBM peak distance in the PDOS. Furthermore, we show the dielectric embedding $GW$ band structure (red) of the CTPA (with the substrate screening effect taken into account implicitly, as shown in Figure \ref{fig:embeddingGW_method}). Here, the positions of the bands are chosen such that their energy differences from the corresponding PBE bands match the self-energies in the dielectric embedding $GW$ calculation. Because the PBE VBM (CBM) energy in Figure \ref{fig:CTPA_pdos_bandstructure}(b) matches the $E_{\rm VBM}-E_{\rm F}$ ($E_{\rm CBM}-E_{\rm F}$) at the interface, the embedding $GW$ VBM (CBM) energy therefore reflects our prediction of the corresponding energy level alignment at the interface using a $GW$-level method.

Our dielectric embedding $GW$ results agree quantitatively with scanning tunneling spectroscopy measurement \cite{SGAY17}. The dielectric embedding $GW$ approach places the CTPA VBM at 1.67 eV below the $E_{\rm F}$ of the interface while Ref. \cite{SGAY17} reports an energy of 1.63 eV. Moreover, our dielectric embedding $GW$ places the CTPA CBM at 1.14 eV above the $E_{\rm F}$, which falls in between the CBM onset (0.82 eV) and the CBM peak (1.51 eV) reported in Ref. \cite{SGAY17}. At the $GW$ level of theory, the fundamental band gap of CTPA reduces from 4.21 eV in the freestanding phase to 2.81 eV at the CTPA/Au(111) interface. It is also noteworthy that the occupied flat bands are below the two Dirac bands in the dielectric embedding $GW$, qualitatively similar to the $GW$ results of the freestanding CTPA.

In addition to comparison with experiment, we benchmark our dielectric embedding $GW$ results to two other $GW$-based computational methods, in a similar fashion as we did in Ref. \cite{L20}: (i) projection $GW$ \cite{CTQ17, TDQB11}, where one calculates $\braket{\phi_i^{\rm COF}|\Sigma[G^{\rm tot}W^{\rm tot}]|\phi_i^{\rm COF}}$, with $\phi_i^{\rm COF}$ a COF orbital (VBM or CBM) and $G^{\rm tot}$ the Green's function of the CTPA/Au(111) interface; (ii) substrate screening $GW$ \cite{LJLN19}, where one calculates $\braket{\phi_i^{\rm tot}|\Sigma[G^{\rm tot}W^{\rm tot}]|\phi_i^{\rm tot}}$, with $\phi_i^{\rm tot}$ an interface orbital representing CTPA VBM or CBM resonance. To reduce the computational cost, such comparisons are performed using a lower cutoff for the dielectric function: 3.3 Ry, corresponding to 2000 bands for the embedding $GW$ and 6000 bands for both the projection $GW$ and the substrate screening $GW$. This parameter has been tested to yield converged band gaps, although the absolute energies of each level (as measured with respect to the vacuum) might not be fully converged. Both the dielectric embedding $GW$ and the projection $GW$ yield a CTPA gap of 2.7 eV on Au(111), verifying the accuracy of our dielectric embedding $GW$ approach, which truncates the substrate polarizability to the adsorbate simulation cell. The substrate screening $GW$ yields a CTPA gap of 2.3 eV on Au(111), and the difference originates from the fact that $\braket{\phi_i^{\rm tot}|\phi_i^{\rm COF}}$ deviates from unity, especially for the CBM.

Lastly, to quantitatively characterize the shape of the kagome bands, we compare the effective masses ($m^*$) at the $\Gamma$ point, from PBE, $GW$, and dielectric embedding $GW$. Given the dispersionless nature of the CBM ($m^*\to\infty$), we only calculate the $m^*$ at VBM, along both $\Gamma \to \mbox{K}$ and $\Gamma \to \mbox{M}$ directions. Along $\Gamma \to \mbox{K}$, the PBE, $GW$, and dielectric embedding $GW$ results are 0.47, 0.46, and 0.46, respectively, in the unit of $m_0$, the electron mass; along $\Gamma \to \mbox{M}$, these three values are 0.46, 0.43, and 0.43, respectively. Our PBE results are consistent with Ref. \cite{JH19}. We conclude that the bands are slightly more dispersive in $GW$ and embedding $GW$ than PBE, and the substrate does not significantly modulate the dispersion of the bands for this material.

\subsection{CTP and its adsorption on Au(111)}
If one replaces the nitrogen atoms in the center of the triangulenes in CTPA [see Figure \ref{fig:cof}(a)] by carbon, one obtains the structure of CTP. The material is reminiscent of graphene \cite{JH19}, being a semi-metal and featuring Dirac bands crossing at the K point at the Fermi level. On the other hand, CTPA can be considered as the analogue of 2D nitrogen in the honeycomb lattice, such that the Fermi level shifts up to the top of the two Dirac bands \cite{JH19}.

\begin{figure}[h]
\includegraphics[width=3.3in]{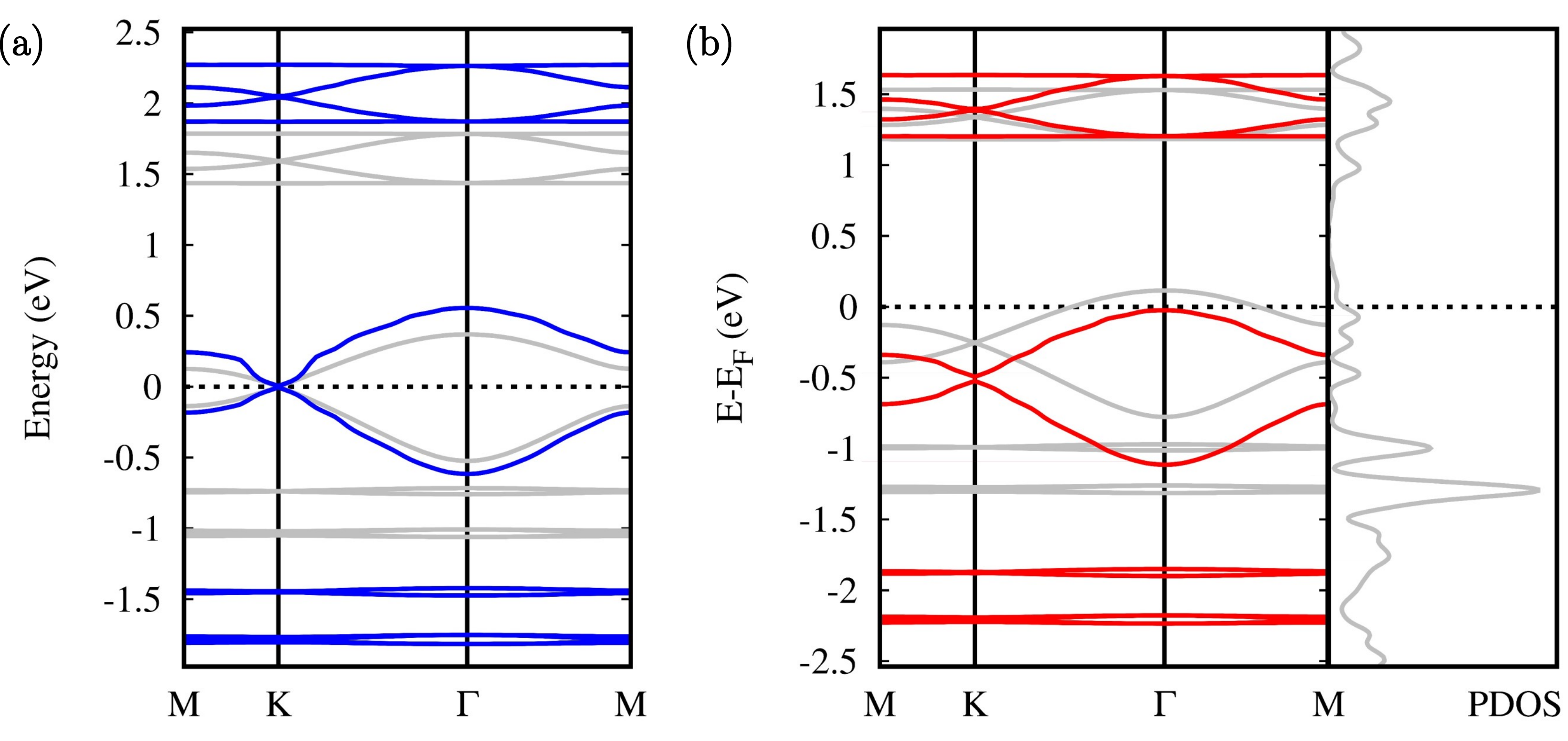}
\caption{(a) PBE (gray) and $GW$ (blue) band structures of the freestanding CTP; (b) PBE (gray) and dielectric embedding $GW$ (red) band structures of the CTP, with band energies shifted such that the energy of PBE Dirac point, $E_{\rm D}$, matches the corresponding energy level alignment at the CTP/Au(111) interface, i.e., $E_{\rm D}-E_{\rm F}$, where $E_{\rm F}$ is the Fermi level of the interface. The PBE PDOS of the CTP/Au(111) interface onto the CTP is shown for comparison. The positions of the embedding $GW$ bands are chosen such that their energy differences from the corresponding PBE bands match the self-energies in the dielectric embedding $GW$ calculation. In this way, the Dirac point in the dielectric embedding $GW$ reflects the prediction of $E_{\rm D}-E_{\rm F}$ at the CTP/Au(111) interface using this approach.}
\label{fig:CTP_pdos_bandstructure}
\end{figure}

Figure \ref{fig:CTP_pdos_bandstructure}(a) shows our calculations of the band structure of the freestanding CTP at both PBE (gray) and $GW$ (blue) levels. Here, all the bands are shifted with respect to the Dirac point that is set as $E=0$. Same as CTPA, the $GW$ and PBE band structures share the same qualitative feature: a flat CBM and the Dirac cone structure around the Fermi level. $GW$ shows a widening of the Dirac bands than PBE, as can be seen at the M and $\Gamma$ points. Also same as CTPA, we see that the occupied flat bands are shifted further away from the two Dirac bands in $GW$. Notably, close to the Dirac point at K, we notice a ``kink" making the $GW$ band structure deviate from the linear dispersion. This phenomenon is reminiscent of that in graphene, which was carefully discussed in Ref. \cite{TGRL08} and was also observed in experiment \cite{BOSH07}. Such a ``link'' affects the way we compute the Fermi velocity, as we discuss below.

Figure \ref{fig:CTP_pdos_bandstructure}(b) shows the electronic structure at the CTP/Au(111) interface, which is a semi-metal/metal interface. We first show the PBE PDOS of the interface onto CTP, with $E=0$ defined as the $E_{\rm F}$ of the interface. Then we show the PBE band structure (gray) of the freestanding CTP [the same one as in Figure \ref{fig:CTP_pdos_bandstructure}(a)] again, but shift the bands such that the energy of PBE Dirac point, $E_{\rm D}$, matches $E_{\rm D}-E_{\rm F}$ at the interface. One can then correlate the band structure with the PDOS. Furthermore, we show the dielectric embedding $GW$ band structure (red) of the CTP, with the substrate screening effect taken into account. Here, the positions of the bands are chosen such that their energy differences from the corresponding PBE bands match the self-energies in the dielectric embedding $GW$ calculation. In this way, the $E_{\rm D}-E_{\rm F}$ (and similar quantities for all bands) in the dielectric embedding $GW$ reflect our prediction of the quasiparticle electronic structure at the interface using a $GW$-level method. We note that the ``kink" close to the Dirac point that we observed in the $GW$ band structure of the freestanding CTP is less pronounced in the embedding $GW$ results. A similar observation has been reported in graphene, where an increase in the electrostatic doping diminishes the ``kink" \cite{AR09}.

\begin{figure}[htp]
\includegraphics[width=3.3in]{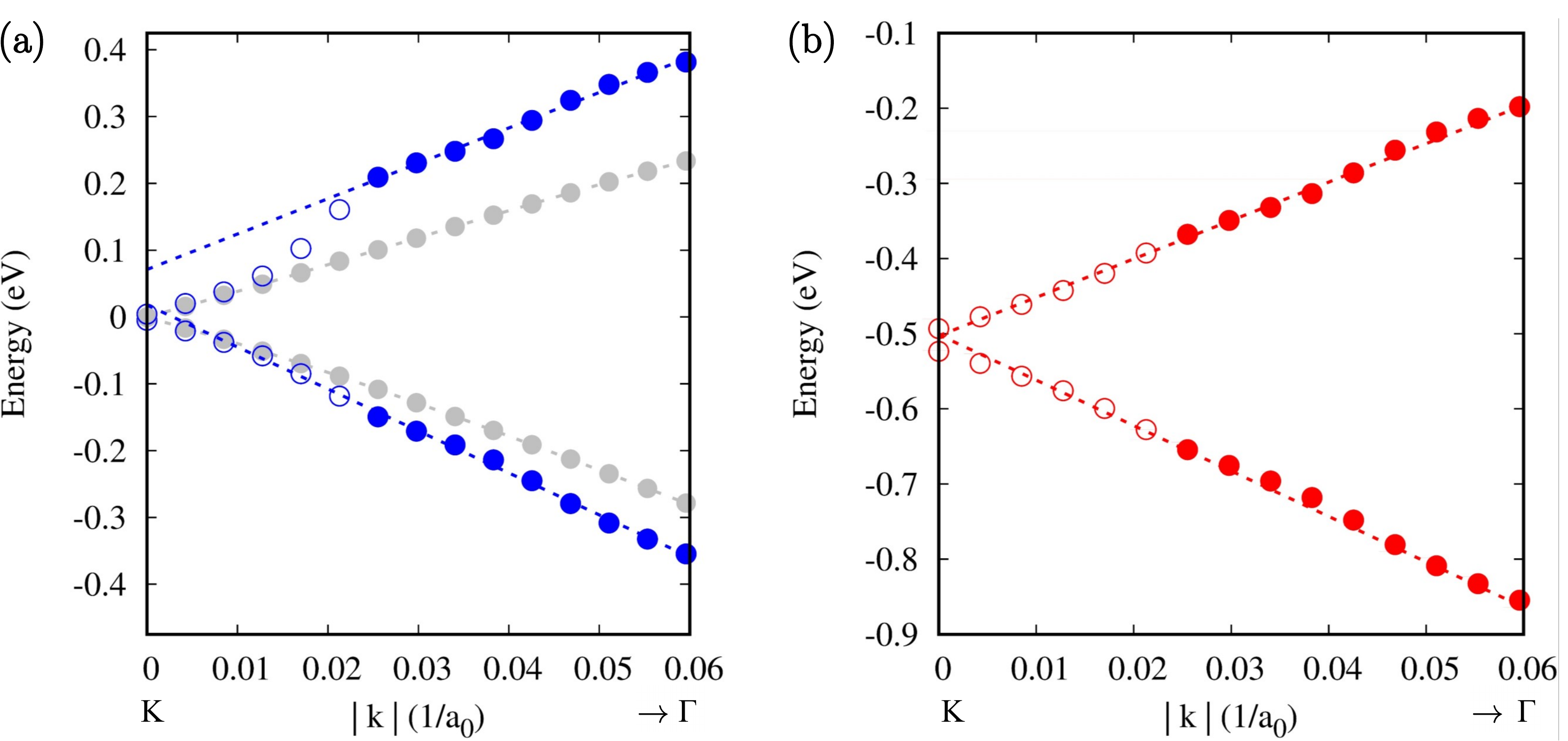}
\caption{Zoomed-in view of (a) PBE (gray circles), $GW$ (both filled and unfilled blue circles), and (b) dielectric embedding $GW$ (both filled and unfilled red circles) band structures near the Dirac point of CTP. In both (a) and (b), to compute the Fermi velocity, a linear fitting (dashed lines) is performed only for the states represented by the filled circles. We use the same convention of $E=0$ as in Figure \ref{fig:CTP_pdos_bandstructure}. $a_0$ is the Bohr radius.}
\label{fig:fermi_velocity}
\end{figure}

Lastly, we compare the Fermi velocities calculated using PBE, $GW$, and dielectric embedding $GW$. Figure \ref{fig:fermi_velocity}(a) shows the zoomed-in view of the PBE (gray dots) and $GW$ (both filled and unfilled blue circles) band structures near the Dirac point (K), and one can clearly see the ``kink'' in the $GW$ band structure. To compute the PBE Fermi velocity, we perform a linear fit (gray dashed lines) using all gray dots shown in Figure \ref{fig:fermi_velocity}(a), resulting in $0.32\times10^{6}$ m/s. To compute the $GW$ Fermi velocity, due to the ``kink'', we perform a linear fit (blue dashed lines) using only the solid blue circles, which is consistent with the practice in Ref. \cite{TGRL08}. This procedure yields Fermi velocities of $0.42\times10^{6}$ m/s and $0.51\times10^{6}$ m/s for the conduction and valence band, respectively. Figure \ref{fig:fermi_velocity}(b) shows the zoomed-in view of the dielectric embedding $GW$ band structure near the Dirac point (both filled and unfilled red circles). One can see that the ``kink'' is not as pronounced as in the $GW$ calculation. To compute the embedding $GW$ Fermi velocity, we again perform a linear fit (red dashed lines) using only the solid red circles, resulting in $0.41\times10^{6}$ m/s and $0.49\times10^{6}$ m/s for the conduction and valence band, respectively.

\subsection{CTP-BN and its adsorption on Au(111)}

If one replaces half of the nitrogen atoms in the centers of the triangulenes in CTPA by boron, one obtains the structure of CTP-BN. Equivalently, it can be obtained by replacing the carbon atoms in the centers of the triangulenes in CTP with boron and nitrogen - in a similar fashion as one obtains hBN from graphene. To the best of our knowledge, CTP-BN has neither been synthesized nor computationally discussed. Here, we use PBE, $GW$, and dielectric embedding $GW$ approaches to explore its electronic structure in the freestanding phase and upon adsorption on Au(111). As we show below, similar to hBN, the nitrogen and boron ``doping'' in CTP-BN opens the gap at the K point compared to CTP, resulting in a wide-gap semiconductor.

\begin{figure}[htp]
\includegraphics[width=3.3in]{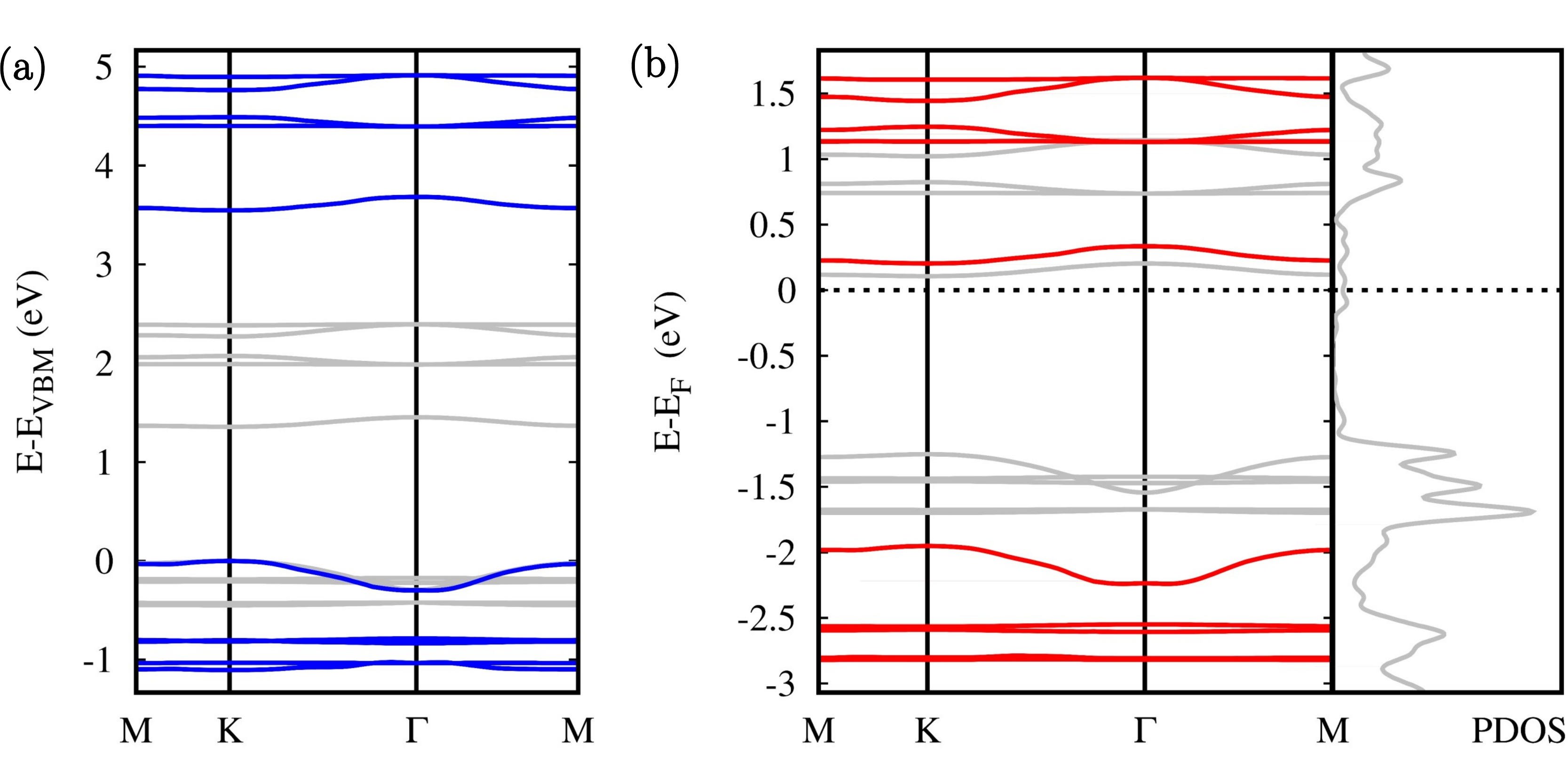}
\caption{(a) PBE (gray) and $GW$ (blue) band structures of the freestanding CTP-BN; (b) PBE (gray) and dielectric embedding $GW$ (red) band structures of the CTP-BN, with band energies shifted such that the value of the PBE VBM matches its energy level alignment at the CTP-BN/Au(111) interface, i.e., $E_{\rm VBM}-E_{\rm F}$. The PBE PDOS of the CTP-BN/Au(111) interface onto the CTP-BN is shown for comparison. The positions of the embedding $GW$ bands are chosen such that their energy differences from the corresponding PBE bands match the self-energies in the dielectric embedding $GW$ calculation. In this way, the embedding $GW$ VBM (CBM) reflects the prediction of $E_{\rm VBM}-E_{\rm F}$ ($E_{\rm CBM}-E_{\rm F}$) at the interface using this approach.}
\label{fig:CTPBN_pdos_bandstructure}
\end{figure}

\begin{table*}[htp]
\caption{Effective masses ($m^*$) of CTP-BN, calculated using PBE, $GW$, and the dielectric embedding $GW$, at both $\Gamma$ and K points, for both CB and VB. All values are in unit of $m_0$. Positive values indicate convex bands that are usually observed at the CBM of a typical semiconductor. Negative values indicate concave bands that are usually observed at the VBM of a typical semiconductor. Due to the complexity of $GW$ and embedding $GW$ valance bands near the $\Gamma$ point, the $m^*$ values are calculated in two regions: values outside (inside) the parentheses are calculated using the states represented by filled circles (diamonds) in Figure \ref{fig:effective_mass}.} \begin{centering}
\begin{tabular}{ c | clcl | cccc }
\hline
\hline
&	\multicolumn{2}{c}{$\Gamma\to\mbox{K}$} & \multicolumn{2}{c|}{$\Gamma\to\mbox{M}$}  & \multicolumn{2}{c}{$\mbox{K}\to\Gamma$} & \multicolumn{2}{c}{$\mbox{K}\to\mbox{M}$} \\
& CB  &  VB  &  CB  &  VB  &CB & VB & CB & VB \\
\hline
PBE   & -0.81 & 0.12 & -0.83 & 0.12 & 1.72  & -0.83 & 1.42 & -0.83 \\
$GW$   & -0.92 & 0.27 (1.06) & -0.91 & 0.25 (-0.46)  & 1.32 & -0.54 & 1.57 & -0.59 \\
embedding $GW$ & -0.94 & 0.27 (-0.31)  & -0.92 & 0.26 (-0.26) & 1.31 & -0.56 & 1.56 & -0.62 \\
\hline
\hline
\end{tabular}
\par\end{centering}
\label{tab:em}
\end{table*}

Figure \ref{fig:CTPBN_pdos_bandstructure}(a) shows the band structure of the freestanding CTP-BN calculated using PBE (gray) and $GW$ (blue), respectively. For comparison, we align the VBM of the PBE and $GW$ band structures at $E=0$. Similar to the above two systems, PBE and $GW$ are qualitatively similar in the features of the kagome bands, with both the VBM and CBM located at the K point. Compared to PBE results, the occupied flat bands are shifted away from the dispersive Dirac bands in $GW$. Notably, at the $\Gamma$ point, PBE predicts one of the flat bands as the top valance band, while $GW$ predicts the dispersive Dirac band as the top valance band. The PBE band gap is calculated to be 1.36 eV, while the $GW$ value is 3.55 eV. 

Figure \ref{fig:CTPBN_pdos_bandstructure}(b) shows the electronic structure at the CTP-BN/Au(111) interface. We first show the PBE PDOS of the interface onto CTP-BN, whith $E=0$ defined as the $E_{\rm F}$ of the interface. Then we show the PBE band structure (gray) of the freestanding CTP-BN [the same one as in Figure \ref{fig:CTPBN_pdos_bandstructure}(a)] again, but shift the energies such that the PBE VBM matches $E_{\rm VBM}-E_{\rm F}$ at the interface. One can then correlate the band structure with the PDOS and find that the PBE CBM-VBM gap is similar for both the freestanding CTP-BN and the CTP-BN adsorbed on Au(111). Furthermore, we show the dielectric embedding $GW$ band structure (red) of the CTP-BN with the substrate screening effect taken into account. Here, the positions of the bands are chosen such that their energy differences from the corresponding PBE bands match the self-energies in the dielectric embedding $GW$ calculation. In this way, the embedding $GW$ VBM (CBM) reflects our prediction of $E_{\rm VBM}-E_{\rm F}$ ($E_{\rm CBM}-E_{\rm F}$) using a $GW$-level method. The dielectric embedding $GW$ yields a gap of 2.16 eV, with VBM 1.95 eV below $E_{\rm F}$ and CBM 0.21 eV above $E_{\rm F}$, both at the K point, decreased from 3.55 eV in the freestanding CTP-BN. 

\begin{figure}[b]
\includegraphics[width=3.3in]{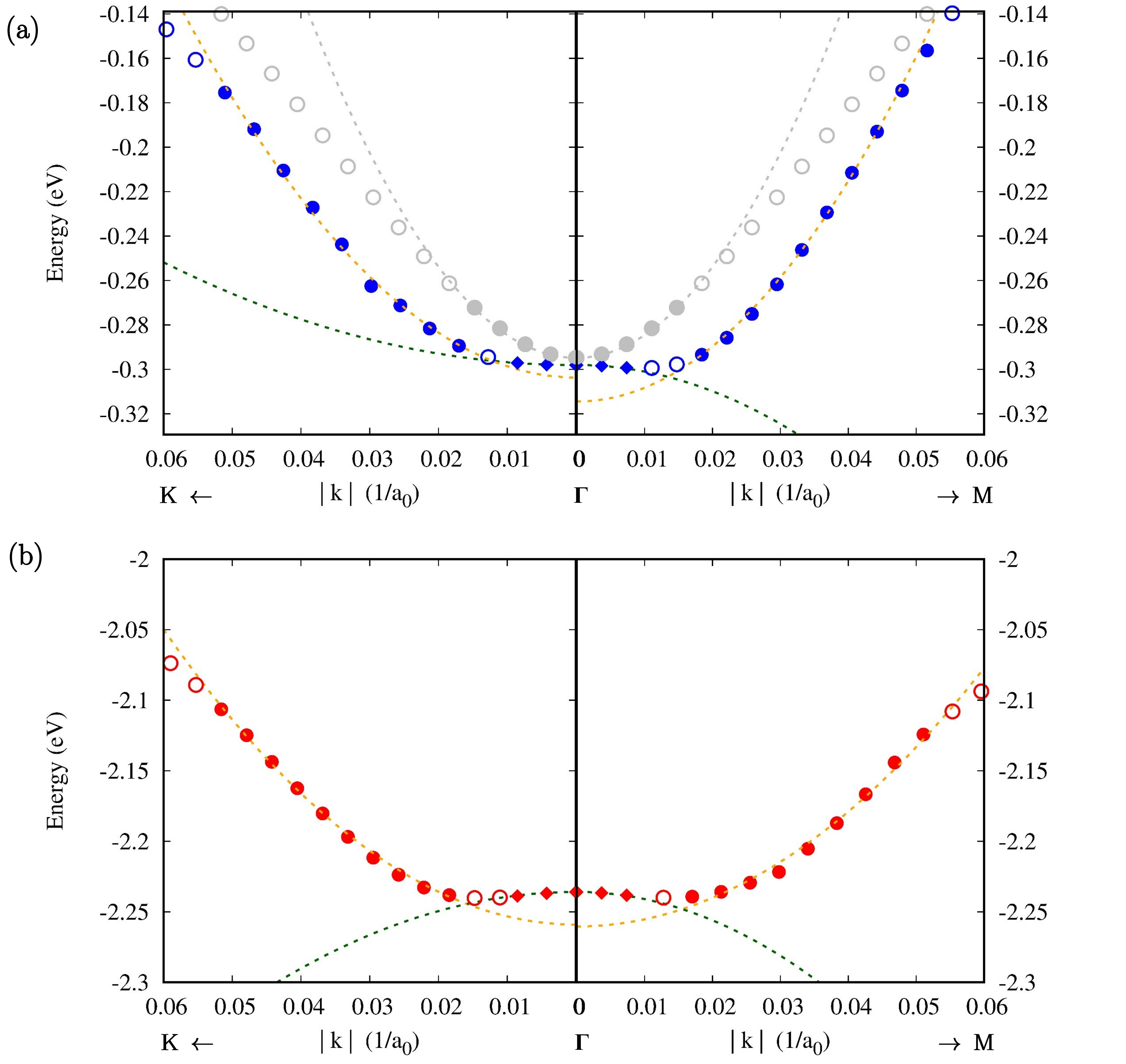}
\caption{Zoomed-in view of (a) PBE (gray symbols), $GW$ (blue symbols), and (b) dielectric embedding $GW$ (red symbols) bands of CTP-BN. Only the occupied dispersive Dirac band near the $\Gamma$ point is shown. In (a), the gray dashed line shows the quadratic fitting of the PBE band. In both (a) and (b), filled diamonds and the green dashed lines (filled circles and the orange dashed lines) show the data points and quadratic fitting of the bands within (outside) the region in close proximity to $\Gamma$. We use the same convention of $E=0$ as in Figure \ref{fig:CTPBN_pdos_bandstructure}.}
\label{fig:effective_mass}
\end{figure}

Lastly, we compare the effective masses calculated from PBE, $GW$, and dielectric embedding $GW$. We show the zoomed-in view of the occupied dispersive Dirac band near the $\Gamma$ point, calculated using PBE (gray symbols) and $GW$ (blue symbols) in Figure \ref{fig:effective_mass}(a), and dielectric embedding $GW$ (red symbols) in Figure \ref{fig:effective_mass}(b). Note that this band is actually not the top valence band at $\Gamma$ in PBE, but is the top band in both $GW$ and dielectric embedding $GW$. The PBE band reaches a minimum at $\Gamma$, qualitatively similar to the lower Dirac band within the valance bands of CTPA (c.f. Figures \ref{fig:CTPA_pdos_bandstructure} and \ref{fig:CTPBN_pdos_bandstructure}). However, $GW$ and dielectric embedding $GW$ bands are bent, reaching a maximum at $\Gamma$, as one can see in the zoomed-in view in Figure \ref{fig:effective_mass}(a)(b). To compute PBE effective masses, we fit the filled gray circles using a quadratic form (gray dashed line). For both $GW$ and dielectric embedding $GW$, we perform the fitting in two regions: one in close proximity to $\Gamma$ (filled blue/red diamonds and the green dashed lines), and one outside this region (filled blue/red circles and the orange dashed lines). The goal behind this strategy is to both characterize the actual curvatures at $\Gamma$ (green dashed lines) and show how $GW$ and dielectric embedding $GW$ modulate the overall shape of PBE bands (orange dashed lines).

The results are summarized in Table \ref{tab:em}. Positive values indicate convex bands that are usually observed at the CBM of a typical semiconductor. Negative values indicate concave bands that are usually observed at the VBM of a typical semiconductor. As we discussed above, the curvatures of the $GW$ and dielectric embedding $GW$ bands are qualitatively different from that of the PBE, for the top valance band in close proximity to the $\Gamma$ point. We therefore list the effective masses calculated in both regions: the values within parentheses are calculated in the region in close proximity to $\Gamma$ (filled blue/red diamonds in Figure \ref{fig:effective_mass}), and the values outside the parentheses are calculated outside this region (filled blue/red circles in Figure \ref{fig:effective_mass}).

\section{Conclusions}\label{conclusion}
In this work, we investigated the quasiparticle electronic structure of heterogeneous interfaces formed between 2D heterotriangulene-based COFs and Au(111), with the aim of understanding the interactions between these 2D COFs and metal substrates. The calculations of such large-scale interfaces were made possible thanks to the newly developed first-principles dielectric embedding $GW$ approach. In particular, we focused on the band gap renormalization, energy level alignment at the interface, and the modulation of the shapes of the COF bands by the substrate. We considered three COFs: CTPA, CTP, and CTP-BN, and showed how the band gaps and the shapes of their kagome bands - as measured by the effective masses and Fermi velocities - vary between the freestanding COFs and those adsorbed on Au(111). We comment that because of the nature of the embedding approach, we only captured the many-body effect (dielectric screening) in the modulation of the shapes of the COF bands by the substrate, and left out the one-body effect (orbital hybridization). Nevertheless, we expect that our approach provides quantitative descriptions of the band gap renormalization and interfacial energy level alignments, as we showed in the explicit comparison against experimental measurements for the CTPA/Au(111) interface. Our results on the CTP/Au(111) and CTP-BN/Au(111) interfaces provide benchmark results for future experiments and calculations. 

\section{Acknowledgements}
Z.-F.L. acknowledges Wayne State University for generous start-up funds, as well as support from the American Chemical Society Petroleum Research Fund (award No. 61117-DNI10). This research used computational resources at the Center for Functional Nanomaterials, which is a U.S. Department of Energy Office of Science Facility, at Brookhaven National Laboratory under contract No. DE-SC0012704. Additional computational resources were provided by the National Energy Research Scientific Computing Center (NERSC), a U.S. Department of Energy Office of Science User Facility operated under contract No. DE-AC02-05CH11231.

\bibliography{lit_COF_Au}
\end{document}